# Bipartite magnetic parent phases in the iron-oxypnictide superconductor

Masatoshi Hiraishi,[1] Soshi Iimura,[2] Kenji M. Kojima,[1,3*] Jun-ichi Yamaura,[4] Haruhiro Hiraka,[1] Kazutaka Ikeda,[1] Ping Miao,[1,3] Yoshihisa Ishikawa,[1] Shuki Torii,[1] Masanori Miyazaki,[1] Ichihiro Yamauchi,[1] Akihiro Koda,[1,3] Kenji Ishii,[5] Masahiro Yoshida,[5,6] Jun'ichiro Mizuki,[6] Ryosuke Kadono,[1,3] Reiji Kumai,[1,3] Takashi Kamiyama,[1,3] Toshiya Otomo,[1,3] Youichi Murakami,[1,3] Satoru Matsuishi,[4] Hideo Hosono[2,4]

[1]Institute of Materials Structure Science, High Energy Accelerator Research Organization (KEK), Tsukuba, Ibaraki 305-0801, Japan

[2]Materials and Structures Laboratory, Tokyo Institute of Technology, Yokohama, Kanagawa 226-8503, Japan.

[3]Department of Materials Structure Science, The Graduate University for Advanced Studies, Tsukuba, Ibaraki 305-0801, Japan

[4]Materials Research Center for Element Strategy, Tokyo Institute of Technology, Yokohama, Kanagawa 226-8503, Japan.

[5]SPring-8, Japan Atomic Energy Agency, Sayo, Hyogo 679-5148, Japan

[6]School of Science and Technology, Kwansei Gakuin University, Sanda, Hyogo 669-1337, Japan

*e-mail: kenji.kojima@kek.jp

**High-temperature (high-$T_c$) superconductivity appears as a consequence of the carrier-doping of an undoped parent compound exhibiting antiferromagnetic order; thereby, ground-state properties of the parent compound are closely relevant to the superconducting state[1,2]. On the basis of the concept, a spin-fluctuation has been addressed as an origin of pairing of the superconducting electrons in cuprates[1]. Whereas, there is growing interest in the pairing mechanism such as an unconventional spin-fluctuation or an advanced orbital-fluctuation due to the characteristic multi-orbital system in iron-pnictides[3-6]. Here, we report the discovery of an antiferromagnetic order as well as a unique structural transition in electron-overdoped**

**LaFeAsO$_{1-x}$H$_x$ ($x \sim 0.5$), whereby another parent phase was uncovered, albeit heavily doped. The unprecedented two-dome superconducting phases observed in this material can be interpreted as a consequence of the carrier-doping starting from the original at $x \sim 0$ and advanced at $x \sim 0.5$ parent phases toward the intermediate region[7]. The bipartite parent phases with distinct physical properties in the second magnetic phase provide us with an interesting example to illustrate the intimate interplay among the magnetic interaction, structural change and orbital degree of freedom in iron-pnictides.**

Iron-pnictides comprise a new class of high-$T_c$ superconductors that has been extensively studied since the discovery of the iron-oxypnictide LaFeAsO (La1111) [8-14]. The superconductivity appears as a result of carrier-doping to the parent compound in place of the magnetic- and structural-ordered state[2,9,15]. A recent study reported an advanced doping method using a hydrogen anion instead of fluorine in La1111 that has surpassed the doping-limit of fluorine, and uncovered the concealed second superconducting phase (SC2) with a higher $T_c$ of 36 K at $x \sim 0.35$, in addition to the first dome (SC1) with the maximum $T_c$ of 26 K at $x \sim 0.1$[7,16]. To investigate the origin of the two SC domes, as well as the origin of the higher $T_c$ in SC2, and to determine whether a certain hidden phase exists beyond the SC2 region, we have performed a multi-probe study in the range of $0.40 \leq x \leq 0.51$ using neutron, muon, and X-ray beams.

Figure 1a shows the neutron powder-diffraction pattern in the non-superconducting specimen with $x = 0.51$. Extra peaks were observed with indices $(1/2, 1/2, n)_{T,M}$ ($n = 0, 1, 2$) in the low-temperature, where the subscripts T and M refer to the tetragonal cell and magnetic peak, respectively; these were determined to be unambiguously magnetic in origin because they were unobservable in the X-ray measurement. The $(1/2, 1/2, 0)_{T,M}$ reflection gradually gains intensity below the magnetic transition temperature $T_N = 89(1)$ K, estimated by the power law in the $x = 0.51$ sample, as shown in Fig. 1b, which indicates a continuous second-order transition. Similar magnetic peaks were also observed below $T_N \sim 76$ K in $x = 0.45$.

The in-plane magnetic structure is illustrated in the inset of Fig. 1b[17,18]. The ordering wavevector $\mathbf{q} = (1/2, 1/2, 0)$ was determined to correspond to the magnetic unit cell of $\sqrt{2}a_T \times \sqrt{2}b_T \times c_T$, describing a tetragonal axis, which is consistent with the orthorhombic nuclear unit cell in the low-temperature. The Fe spins are directed to a diagonal axis of the tetragonal-cell, and form an antiferromagnetic collinear structure; this result reveals an exceptional stripe-type arrangement among the iron-pnictides[19-22]. The spins couple ferromagnetically along the $c$-axis (right inset in Fig. 4). The magnetic moment in $x = 0.51$ at 10 K is estimated to be 1.4 $\mu_B$ per iron atom, which is about four times as large as 0.36 $\mu_B$ in $x = 0$[19,20]. This observation that the long-range magnetic order emerges beyond SC2 is a remarkable finding because the $d$-electrons on the iron atoms should be delocalised by the large amount of carrier-doping, and accordingly, the magnetic interaction is usually expected to be weak.

Figure 2a shows the muon-spin-relaxation (μSR) time spectra in $x = 0.45$. The slow relaxation above 80 K is attributed to a paramagnetic state, whereas, the line-shapes below 80 K evolve into strongly damped oscillations with increasing amplitude on cooling, reflecting the development of the magnetically ordered volume fraction (MVF).

Figure 2b illustrates the diagram of the μSR process. The amplitude and frequency of the μSR signal reflect the MVF and the ordered moment <μ> of the magnetic region, respectively. Figure 2c indicates the temperature dependence of the resultant MVF. In $x = 0.51$, the magnetic ordering appears at $T_N = 92(7)$ K, and then, the MVF becomes entirely toward the zero temperature. As the hydrogen content is reduced, the $T_N$ and the MVF decrease together. Figure 2d shows the MVF and the superconductivity volume fraction (SVF) estimated from the susceptibility measurements[7], clearly indicating an inverse correlation between the MVF and the SVF. The approximate unity in the summation suggests that antiferromagnetic static order and superconductivity coexist in the range $0.40 \leq x \leq 0.45$, despite the absence of the coexistence state in the underdoped region[15]. This phenomenon is known as the coexistence state or the microscopic phase separation of superconductivity and magnetism, which is considered to be the signature of strong electron correlation and an intimate connection between superconductivity and magnetism in high-$T_c$ superconductors[24,25].

Figure 3a shows the X-ray profiles of the (2, 2, 0)$_T$ reflections in $x = 0.45$ and 0.51. Upon cooling, the peak in $x = 0.51$ was broadened moderately, while the peak in $x = 0.45$ exhibited slight broadening. No broadening of the (0, 0, $l$) reflections was observed in $x = 0.51$;

therefore, the experimental findings imply that the tetragonal to orthorhombic (T-O) structural transition emerges obviously in $x = 0.51$.

Figure 3b indicates the temperature dependence of the lattice constants in $x = 0.45$ and 0.51. In $x = 0.51$, the $a_T$-axis length splits in two below the T-O transition of $T_s \sim 95$ K, and the $c_T$-axis length shows an upturn at $T_s$. In $x = 0.45$, although an apparent $a_T$-$b_T$ split does not exist, the $c_T$-axis anomaly appears to be similar to that in $x = 0.51$, which may be the result of an insufficient coherence of $a_T$-$b_T$ splitting in the low-temperature phase.

The resultant temperature dependence of the structural and magnetic order parameters are shown in Fig. 3c in $x = 0.51$. The small gap between the transition temperatures $T_s$ and $T_N$ implies a strong correlation between structure and magnetism in comparison with the large gap of ~20 K in $x = 0$. On the basis of a structural analysis[26], the compound with $x = 0.51$ crystallises below $T_s$ in an orthorhombic *Aem*2 structure without inversion symmetry, in contrast to the universal *Cmme* structure with inversion symmetry observed in the 1111 materials. The Fe atom exhibits the off-centre deformation in the FeAs$_4$ tetrahedron as shown in the inset of Fig. 4. The present result provides us with an intriguing deduction that the unusual structural distortion, along with the loss of the inversion symmetry, which has not been found in large ensemble of the iron-pnictides,[19-22] may profoundly affect the *d*-orbital levels of the iron atom[27,28].

Figure 4 illustrates the phase diagram of LaFeAsO$_{1-x}$H$_x$. The feature of the physical properties in $x = 0.51$ ($x = 0$) can be characterised as follows: the Fe-spin arrangement is peculiar stripe-type (universal stripe-type)[18,19]; the magnetic moment is large, 1.4 $\mu_B$ (small, 0.36 $\mu_B$); the gap between $T_s$ and $T_N$ is ~5 K (~20 K); the structural symmetry is non-centrosymmetric (centrosymmetric); the magnetic and superconducting states coexist (the magnetic and superconducting states are exclusive) ; the behaviour of resistivity: non-Fermi liquid (Fermi liquid)[7]. These findings strongly suggest the magnetic and electronic correlations considerably-developed in the right-hand region of the phase diagram.

We have thus far regarded the 'undoped' antiferromagnet with the structural transition adjacent to the superconducting state as the parent compound in the high-$T_c$ superconductors. Hence, we now reveal that a new uncovered antiferromagnetic phase with the structural transition adhering to the superconducting phase is considered as the 'doped' parent phase. This behaviour is unexpected because magnetic and electronic correlations are widely perceived as being weak in the heavily carrier-doped region. However, the magnetic and electronic correlations in the advanced parent phase at $x \sim 0.5$ are indeed rather strong in comparison to the original parent phase at $x \sim 0$. We think that the multi-band feature such as an orbital-selective Hund's coupling or Mott-transition plays a crucial role so as to drive the emergence of the advanced parent phase, which is presumably stimulated by the half-integer number of the $d^{6.5}$ state in $x = 0.5$[29,30]. In the iron-pnictides, the 'parent compound' not only refers to the undoped material, but also more generally indicates a fingerprint at a certain critical point of the magnetic and electronic correlations.

We now discuss the origin of the two-dome SC phases. The electron-doping to the $d^6$ state of the $x = 0$ parent, i.e. the chemical substitution of $H^-$ to the $O^{2-}$ sites, leads to the SC1 phase. From the viewpoint of the $x \sim 0.5$ parent, the SC2 phase emerges via hole-doping to the $d^{6.5}$ state, by $O^{2-}$ substitution to the $H^-$ site. Consequently, we can definitely state that the two-dome SC phases are generated by carrier-doping, starting from the left- and right-hand parent compounds towards the intermediate region of the phase diagram. Moreover, the $T_c$ valley at $x \sim 0.2$ is interpreted as a unique crossover region in this phase diagram.

The combination of the distinct feature of the structural instability with the loss of inversion symmetry and the localised nature of the $d$-electrons relative to the original parent compound plausibly result in a certain new mechanism of superconductivity as an origin of the significant enhancement of $T_c$ in SC2 relative to SC1, because the ground-state properties in the parent compounds generally persist unabated, even in the superconducting state. Future band structure calculations taking into account the orthorhombic-crystal and magnetic structures would confirm and clarify this novel ordered state.

**Acknowledgements**

We thank K. Yamada for helpful discussion. The neutron, muon, and synchrotron radiation experiments were performed at J-PARC (BL08-SuperHRPD, BL21-NOVA, Muon D1), PSI, KEK-PF (BL-8A/8B), SPring-8 with the approval of JASRI (BL11XU) (Proposal Nos. 2013S2-002, 2009S05, 2013A3502). This work was supported by MEXT Elements Strategy Initiative to Form Core Research Center.

**References**

. 1. Lee, A. P., Nagaosa, N, & Wen, X-G. Doping a Mott insulator: Physics of high-temperature superconductivity. *Rev. Mod. Phys.* **78**, 17-85 (2006).

. 2. Paglione, J. & Greene, R. L. High-temperature superconductivity in iron-based materials. *Nature Phys.* **6**, 645-658 (2010).

. 3. Singh, D. J. & Du, M.-H. Density functional study of LaFeAsO$_{1-x}$F$_x$: A low carrier density superconductor near itinerant magnetism. *Phys. Rev. Lett.* **100**, 237003 (2008).

. 4. Kuroki, K. *et al.* Unconventional pairing originating from disconnected Fermi surfaces of superconducting LaFeAsO$_{1-x}$F$_x$. *Phys. Rev. Lett.* **101,** 087004 (2008).

. 5. Mazin, I.*,* Singh, D. J., Johannes, M. D. & Du, M. H. Unconventional superconductivity with a sign reversal in the order parameter of LaFeAsO$_{1-x}$F$_x$. *Phys. Rev. Lett.* **101**, 057003 (2008).

. 6. Kontani, H. & Onari, S. Orbital-fluctuation-mediated superconductivity in iron pnictides: Analysis of the five-orbital Hubbard-Holstein model. *Phys. Rev. Lett.* **104**, 157001 (2010).

. 7. Iimura, S. *et al.* Two-dome structure in electron-doped iron arsenide superconductors. *Nature Commun.* **3**, 943 (2012).

. 8. Kamihara, Y. *et al.* Iron-based layered superconductor: LaOFeP. *J. Am. Chem. Soc.* **128**, 10012-10013 (2006).

9. Kamihara, Y., Watanabe, T., Hirano, M. & Hosono, H. Iron-based layered superconductor La[$O_{1-x}F_x$]FeAs ($x$ = 0.05 - 0.12) with $T_c$ = 26 K. *J. Am. Chem. Soc.* **130**, 3296-3297 (2008).

10. Chen, X. H. *et al.* Superconductivity at 43 K in SmFeAsO$_{1-x}$F$_x$. *Nature* **453**, 761-762 (2008).

11. Takahashi, H. *et al.* Superconductivity at 43 K in an iron-based layered compound LaO$_{1-x}$F$_x$FeAs. *Nature* **453**, 376-378 (2008).

12. Rotter, M., Tegel, M. & Johrendt, D. Superconductivity at 38 K in the iron arsenide (Ba$_{1-x}$K$_x$)Fe$_2$As$_2$. *Phys. Rev. Lett.* **101**, 107006 (2008).

13. Yildirim, T. Origin of the 150-K anomaly in LaFeAsO: Competing antiferromagnetic interactions, frustration, and a structural phase transition. *Phys. Rev. Lett.* **101**, 057010 (2008).

14. Ma, F. & Lu, Z.-Y. Iron-based layered compound LaFeAsO is an antiferromagnetic semimetal. *Phys. Rev. B* **78**, 033111 (2008).

15. Luetkens, H. *et al.* The electronic phase diagram of the LaO$_{1-x}$F$_x$FeAs superconductor. *Nature Mater.* **8**, 305-309 (2009).

16. Matsuishi, S. *et al.* Structural analysis and superconductivity of CeFeAsO$_{1-x}$H$_x$. *Phys. Rev. B* **85**, 014514 (2012).

17. Rodriguez-Carvajal, J. Recent advances in magnetic structure determination by neutron powder diffraction. *Physica B* **192**, 55-69 (1993).

18. Larson, A. C. & Von Dreele, R. B. General structure analysis system (GSAS). *Los Alamos National Laboratory Report LAUR.* 86-748 (2000).

19. de la Cruz, C. *et al.* Magnetic order close to superconductivity in the iron-based layered LaO$_{1-x}$F$_x$FeAs systems. *Nature* **453**, 899-902 (2008).

20. Lynn, J. W. & Dai, P. Neutron studies of the iron-based family of high $T_C$ magnetic superconductors. *Physica C* **469**, 469-476 (2009).

21. Huang, Q. *et al.* Neutron-diffraction measurements of magnetic order and a structural transition in the parent $BaFe_2As_2$ compound of FeAs-based high-temperature superconductors. *Phys. Rev. Lett.* **101**, 257003 (2008).

22. Lumsden, M. D. & Christianson, A. D. Magnetism in Fe-based superconductors. *J. Phys. Condens. Matter* **22**, 203203 (2010).

23. Klauss, H. H. *et al.* Commensurate spin density wave in LaFeAsO: A local probe study. *Phys. Rev. Lett.* **101**, 077005 (2008).

24. Uemura, Y. J. Superconductivity: Commonalities in phase and mode. *Nature Mater.* **8**, 253-255 (2009).

25. Takeshita, S. *et al.* Insular superconductivity in a Co-doped iron pnictide $CaFe_{1-x}Co_xAsF$. *Phys. Rev. Lett.* **103**, 027002 (2009).

26. Izumi, F. & Momma, K. Three-dimensional visualization in powder diffraction. *Solid State Phenom.* **130**, 15-20 (2007).

27. Lee, C.-C., Yin, W.-G. & Ku., W. Ferro-orbital order and strong magnetic anisotropy in the parent compounds of iron-pnictide superconductors. *Phys. Rev. Lett.* **103**, 267001 (2009).

28. Lv, W., Wu, J. & Phillips., P. Orbital ordering induces structural phase transition and the resistivity anomaly in iron pnictides. *Phys. Rev. B* **80**, 224506 (2009).

29. de' Medici, L., Hassan, S. R., Capone, M. & Dai, X. Orbital-selective Mott transition out of band degeneracy lifting. *Phys. Rev. Lett.* **102**, 126401 (2009).

30. Lanatà, N. *et al*. Orbital selectivity in Hund's metals: The iron chalcogenides. *Phys. Rev. B* **87**, 045122 (2013).

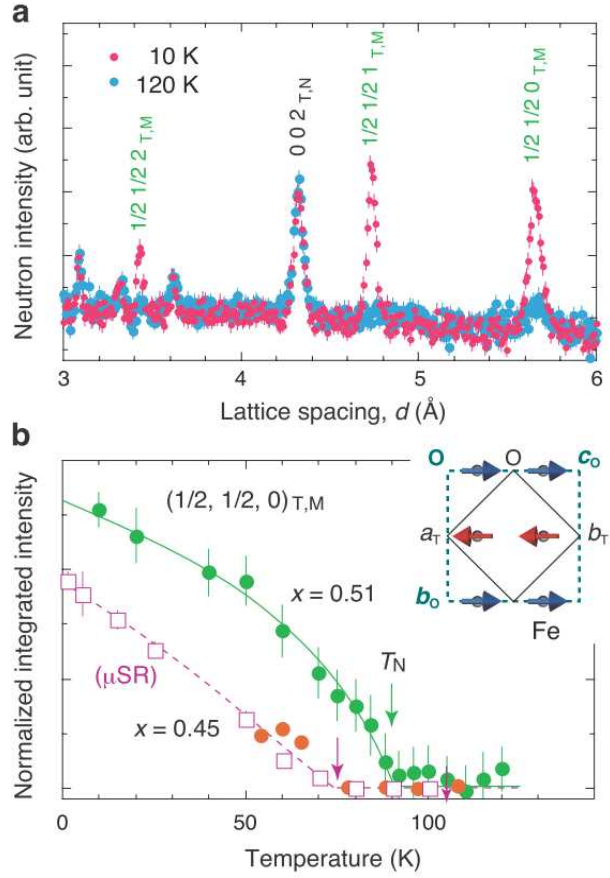

**Figure. 1 | Pulsed neutron diffraction measurements of LaFeAsO$_{1-x}$H$_x$.** We prepared ~1 g powder samples as described in Ref. 7. The powder diffraction measurements were performed on the low-angle detector bank of the Super HRPD at J-PARC for $x = 0.51$ and on the NOVA at J-PARC for $x = 0.45$. **a**, The magnetic reflections of $(1/2, 1/2, n)_{T,M}$ ($n = 0, 1, 2$) were observed clearly at 10 K, which disappears at 120 K. **b**, The temperature dependence of the integrated intensity of $(1/2, 1/2, 0)_{T,M}$ in $x = 0.45$ (orange) and 0.51 (green), normalised to the nuclear $(0, 0, 2)_{T,N}$ reflection. The data in $x = 0.51$ was fitted to the power law $(T_N - T)^{2\beta}$, resulting in $T_N = 89(1)$ K and $2\beta = 0.58(5)$. The squared order parameter for $x = 0.45$ determined by μSR [ $\propto$ (muon frequency)$^2 \times$ (magnetic volume fraction)] is plotted (purple) after scaling to neutron data. The in-plane configuration of the Fe spin, determined by the Rietveld method[17,18], for $x = 0.45$ as well as 0.51 is illustrated in the inset with the tetragonal unit cell $a_T \times b_T$ (solid line) it the high-temperature, and with the magnetic unit cell of $\sqrt{2}a_T \times \sqrt{2}b_T$ (broken line) in the low-temperature.

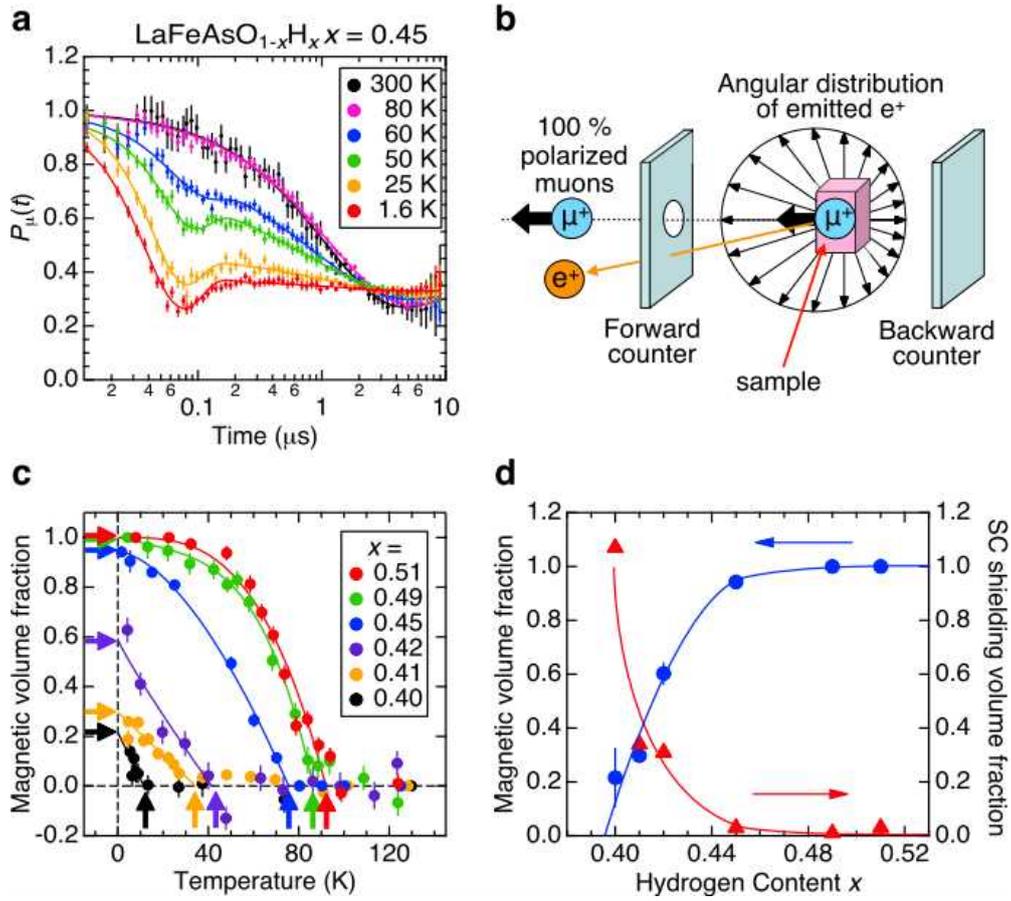

**Figure 2 | Zero-field muon spin relaxation (µSR) measurements. a,** Time spectra in $x =$ 0.45, taken at PSI. The present data exhibit significant damping in contrast to a clear muon spin precession in $x = 0$ (ref. 23), indicating an inhomogeneity in the local magnetic fields. **b,** The diagram of µSR process. The magnetic volume fraction (MVF) is estimated from the amplitude of the time-dependent asymmetry in the positron counts between the forward and backward counters. **c,** The temperature dependence of MVF for samples in the range of $x =$ 0.40 - 0.51, conducted at J-PARC MUSE and PSI. **d,** Hydrogen concentration dependence of the MVF (blue) and the superconductivity volume fraction (red). The inverse correlation suggests a microscopic phase separation between the magnetic and superconducting regions.

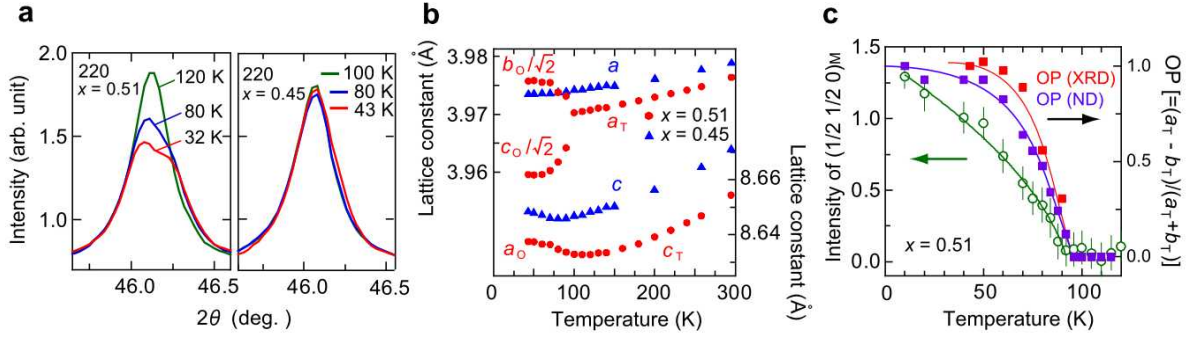

**Figure 3 | Synchrotron X-ray diffraction measurements.** The powder diffraction measurements for samples with $x$ = 0.42, 0.45, 0.49, and 0.51 were performed on the beamlines of 8A/8B at KEK-PF. **a,** Representative profiles of the $(2, 2, 0)_T$ reflections for $x$ = 0.45 and 0.51 with the wavelength of $\lambda$ = 1.0993 Å. A peak broadening was observed in $x$ = 0.51 below $T_s \sim$ 95 K, while there was slight broadening in $x$ = 0.45. **b,** The temperature dependence of the lattice constants for $x$ = 0.45 and 0.51. In $x$ = 0.51, the $a_T$- and $c_T$-axes indicate the split of the length and the upturn with the T-O transition of $T_s \sim$ 95 K, respectively. In the orthorhombic phase, the values of $b_O$ and $c_O$ are divided by $\sqrt{2}$. The cell setting changes as $a_O = c_T$, $b_O = a_T + b_T$ and $c_O = -a_T + b_T$ (see Fig. 1b inset). The orthorhombicity, $\Delta a_T / a_T$ = 0.41%, is smaller than that of the other iron-pnictides (0.49%, LaFeAsO; 0.73%, $BaFe_2As_2$)[18,20]. **c,** The temperature dependence of the structural order parameter (OP) determined by neutron and X-ray diffraction, and the magnetic intensity of the $(1/2, 1/2, 0)_{T,M}$ reflection in $x$ = 0.51.

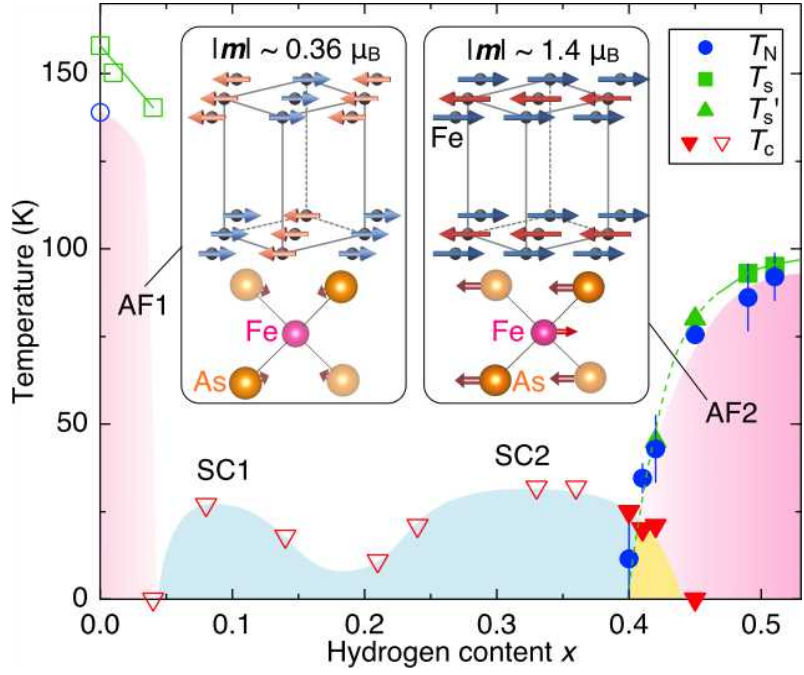

**Figure 4 | Magnetic, structural and superconducting phase diagram of LaFeAsO$_{1-x}$H$_x$.** The original parent compound in $x = 0$ exhibits a structural transition at $T_s = 155$ K, followed by an antiferromagnetic state (AF1) at $T_N = 137$ K[19]. With increasing $x$, two superconductivity domes appear: $0.05 \leq x \leq 0.20$ (SC1) with $T_{c, max} = 26$ K, and $0.20 \leq x \leq 0.42$ (SC2) with $T_{c, max} = 36$ K[7]. Eventually, another antiferromagnetic phase (AF2) appears at $0.40 \leq x \leq 0.51$. In the advanced parent compound at $x = 0.51$, the structural and magnetic transitions occur at $T_s \sim 95$ K and $T_N = 89$ K, respectively. $T_s$' indicates the $c$-axis upturn temperature observed in X-ray measurements. The filled and open marks are obtained from the present and the previous results, respectively[7]. The magnetic structures of AF1 (left) and AF2 (right) are illustrated with their magnetic moments, where the solid lines represent the tetragonal cell. The displacements of the Fe and As atoms across the structural transitions are schematically described by the arrows on the FeAs$_4$ tetrahedra from the view of the orthorhombic long-axis, in which the Fe and As atoms move by 0.07 Å (0 Å) and 0.06 Å (0.01 Å) in $x = 0.51$ ($x = 0$), respectively.